\documentclass[a4paper]{article}
\usepackage{amsmath,epsfig}

\newcommand{\nn}{\nonumber}

\begin{document}

\noindent
{\bf\Large \textsf{Minimax determination of the energy spectrum of \\
    the Dirac equation in a Schwarzschild  background }} 

\vspace{0.2cm}

\noindent 
Alejandro C\'{a}ceres\footnote{ac293@mrao.cam.ac.uk} and
Chris Doran\footnote{c.doran@mrao.cam.ac.uk}

\vspace{0.2cm}

\noindent 
Astrophysics Group, Cavendish Laboratory, Madingley Road, \\
Cambridge CB3 0HE, UK.

\vspace{0.4cm}

\begin{center}

\begin{abstract}
We calculate the bound-state energy spectrum of the Dirac Equation in
a Schwarzschild black hole background using a minimax variational
method.  Our method extends that of Talman~\cite{talman} to the case
of non-Hermitian interactions, such as a black hole.  The trial
function is expressed in terms of a basis set that takes into account
both the Hermitian limit of the interaction in the non-relativistic
approximation, and the general behaviour of the solutions at the
origin, the horizon and infinity.  Using this trial function an
approximation to the full complex energy bound-state spectrum is
computed.  We study the behaviour of the method as the coupling
constant of the interaction is increased, which increases both the
relativistic effects and the size of the non-Hermitian part of the
interaction. Finally we confirm that the method follows the expected
Hylleraas--Undheim behaviour.
\end{abstract}

\vspace{0.2cm}


\end{center}

\section{Introduction}

Variational methods are an important practical technique for
determining the eigenvalues of Hamiltonian operators. The methods
aim to provide an approximation to the energy eigenvalues by the
variation of the Rayleigh quotient
\begin{equation}
E_\phi = \frac{(\phi, {\bf H}\,\phi)}{(\phi, \phi)} 
\label{Rq}
\end{equation}
for a continuous set of trial functions $\phi=\phi(a)$. In
non-relativistic quantum mechanics the energy spectrum is bounded
from bellow and a simple minimization of $E_\phi$ with respect to
$a$ leads to an upper bound to the ground state energy.
Furthermore, if $\phi$ is expressed on a finite-basis expansion,
then the eigenvalues of $H$ on this basis are approximations to
the higher energy eigenvalues.  This is known as the
Hylleraas--Undheim theorem~\cite{hylleraas,mac}.

For the Dirac equation the picture is more complicated as the spectrum
of the Dirac Hamiltonian is not bounded from below.  That is, the
Hamiltonian allows states with positive and negative energy.
Therefore, a direct minimization of the parameter $a$ takes $E_\phi$
to $-\infty$. This is known as variational collapse \cite{sch}. Many
different techniques have been devised to avoid this problem (for a
review of the field, see \cite{dol}).  Drake and Goldman~\cite{gol}
showed that for a well chosen basis the variational approach for the
hydrogen atom leads to sensible results. They found that the energy
eigenvalues split into one positive and one negative set. The positive
set give the bound-state energies, while the negative energies are
smaller than minus the electron's rest mass.  Talman \cite{talman}
explained how the finite basis expansion of the large and small parts
of the trial solution provide the positive and negative eigen-energies
respectively.  Talman also addressed Grant's argument \cite{grant}
that the variational collapse is caused by the incompleteness of the
small-component basis.  Talman discussed how the ground-state energy
of a truncated small-component basis is a lower bound for the energy
of a complete basis. It also turns out that this complete basis energy
is an upper-bound for the exact energy. Dolbeault \emph{et al.}
\cite{dol} have proved that equation (\ref{minmax}) below can be
derived from a general variational principle, and have stressed that
the difficulty in Talman's approach is to determine for which class of
Hamiltonians it is valid. In particular, they have proved that the
approach is valid for Hermitian Hamiltonians with potentials of the
form $|r|^{-\beta}$, where $\beta \in (0,1)$, which evidently include
the Coulomb interaction.

Recently the bound-state spectrum of an electron in a Schwarzschild
black hole back ground has been determined~\cite{las}.  One of its
main features is that the energy eigenvalues are complex, since the
black hole interaction is non-Hermitian. The imaginary part of the
energy eigenvalues gives the decay rate with time of the corresponding
eigenstate.  Furthermore, the state spectrum resembles that of the
hydrogen atom for a small coupling constant. As the coupling constant
is reduced, the imaginary part of the energy tends to zero as the
interaction tends to one of Coulomb type.  This offers a starting point
to explore the variational methods for non-Hermitian Hamiltonians.  In
this parer we thus implement a minimax method based on Talman's
approach for the black hole interaction. This method allows us to
calculate the full complex energy eigenvalues and explore its behaviour
with an increasing coupling constant.  The approach developed here
should have applications in a range of fields where one is forced to
deal with non-Hermitian interactions, including the important subject
of open systems.

\section{Talman's minimax method}

Talman's variational formulation for the energy eigenvalues follows
from the decomposition of the Dirac wavefunction into its large and
small parts. Following Grant \cite{grant}, the two-component radial
Dirac equation can be written as
\begin{equation}
{\bf H}
\begin{pmatrix}
g \\ f
\end{pmatrix}
=
\begin{pmatrix}
1 + \frac{1}{mc^2} H_1(r) & \frac{\hbar}{mc}
\frac{k-1}{r}-\frac{\hbar}{mc}\partial_r \\
\frac{\hbar}{mc}
\frac{k+1}{r}+\frac{\hbar}{mc}\partial_r &-1 + \frac{1}{mc^2}
H_1(r)
\end{pmatrix}
\begin{pmatrix}
g \\ f
\end{pmatrix}
=
\frac{E}{mc^2} 
\begin{pmatrix}
g \\ f
\end{pmatrix}
\label{Dirac1}
\end{equation}
where $H_1(r)$ is the interaction term. The ground state energy
can be obtained from the variation of the correct eigenfunction
$\phi$, determined by $f$ and $g$;
\begin{equation}
E=\mbox{min}_g \left[ \mbox{max}_f\,\,\frac{(\phi, {\bf
H}\,\phi)}{(\phi, \phi)} \right],
\label{minmax}
\end{equation}
where $\phi$ is the two-component radial spinor
\begin{equation}
\phi(r)= 
\begin{pmatrix}
g(r) \\ f(r)
\end{pmatrix} .
\end{equation}
If $g$ is an arbitrary function, the maximum over $f$ yields an
upper-bound to the exact ground energy. The trial function $\phi$ can
be expanded in a finite function basis in the form
\begin{equation}
\phi(r)=\sum_{j=1}^{n} 
\begin{pmatrix}
b^j g_j(r)\\
a^j f_j(r)
\end{pmatrix}.
\label{eqn_trialgen}
\end{equation}
After a substitution of this function into equation
(\ref{minmax}), the minimax approximation is reduced to minimize
$E$ for each $b^j$ and maximize it for each $a^j$. If we then
consider vectors of the form
\begin{equation}
\alpha=(b_1...b_n,a_1,...a_n)^t,
\end{equation}
then the minimax variation of $E$ leads to a generalized eigenvalue
problem. This can be written
\begin{equation}
 {\bf h}\, \alpha = E \, {\bf O}\,\alpha.
\label{eqn_genevalue}
\end{equation}
Here ${\bf h} $ is a $(2n \times 2n)$ matrix
\begin{equation}
 {\bf h}  =
\begin{pmatrix}
{\bf h11}_{n \times n} & {\bf h12}_{n \times n}\\
{\bf h21}_{n \times n} & {\bf h22}_{n \times n}
\end{pmatrix}
\end{equation}
where the $(n \times n)$ matrices are given by
\begin{align}
 {\bf h11}_{lj} & = \int_{0}^{\infty} g_l^*\, {\bf H}_{11}\, g_j\, r^2
 dr\\
{\bf h12}_{lj}  &= \int_{0}^{\infty} g_l^*\, {\bf H}_{12}\, f_j\,
r^2
 dr\\
 {\bf h21}_{lj}  &= \int_{0}^{\infty} f_l^*\, {\bf H}_{21}\, g_j\, r^2
 dr\\
 {\bf h22}_{lj}  &= \int_{0}^{\infty} f_l^*\, {\bf H}_{22}\, f_j\, r^2
 dr.
\end{align}
The matrix ${\bf O}$ is also a $(2n \times 2n)$ matrix given by the
overlap integrals of the $g_l$ and $f_j$.  Its first $(n \times n)$
sub-matrix is
\begin{equation}
 {\bf O11}_{lj}  = \int_{0}^{\infty} g_l^*\,  g_j\, r^2
 dr,
\end{equation}
with ${\bf O22}$ containing the equivalent $f_j$ integrals, and all
other terms vanishing.

Talman shows that for the Coulomb interaction the positive eigenvalues
in equation (\ref{eqn_genevalue}) tend to the first energy eigenvalues
as $n$ increases, following a Hylleraas--Undheim behaviour.

\section{Black hole interaction}

The interaction $H_1$ for a black hole is given by \cite{las2}
\begin{equation}
\frac{1}{mc^2} H_1(r)=i \frac{\hbar}{mc} 
\sqrt{\frac{\hbar}{m c r} \frac{2GMm}{c\hbar} }
(\partial_r +\frac{3}{4r}),
\label{int}
\end{equation}
where the dimensional constants have been arranged into a convenient
 form.  If we measure the radial distance in units of the Compton wavelength
 ($mc/\hbar$) we are left with the dimensionless coupling constant
\begin{equation}
\alpha=\frac{GMm}{c\hbar}
\end{equation}
which gives the magnitude of the interaction.  Using the interaction
(\ref{int}), we can now write the system of differential equations
(\ref{Dirac1}) in a more convenient form.  With the change of
variables
\begin{equation}
\varpi_1=\frac{f}{r} \qquad \mbox{and} \qquad
\varpi_2=\frac{g}{r},
\end{equation}
the equations can be written as
\begin{equation}
  \begin{pmatrix}
       \partial_r \varpi_1\\
       \partial_r \varpi_2
  \end{pmatrix}
=  {\bf C }(r) 
  \begin{pmatrix}
       \varpi_1\\
       \varpi_2
  \end{pmatrix},  
\label{eq_Cform}
\end{equation}
where ${\bf C}$ is the matrix
\begin{equation}
{\bf C }(r) = \frac{1}{r-2\alpha}
\begin{pmatrix}
        k-\frac{\alpha}{2r}+ i\sqrt{2\alpha r}(1+E)        &
        i(-k-\frac{1}{4})\sqrt{\frac{2\alpha}{r}} + r(1-E)\\
        i(-k+\frac{1}{4})\sqrt{\frac{2\alpha}{r}} + r(1+E) &
        -k-\frac{\alpha}{2r}- i\sqrt{2\alpha r}(1-E)
  \end{pmatrix}
\end{equation}
and we measure $r$ and $E$ in units of ($mc/\hbar$) and ($mc^2$)
respectively.  To simplify some of the later expressions we set the
Compton wavelength $mc/\hbar$ equal to 1.

Special attention must be given to the singularities of equation
(\ref{eq_Cform}). The equation has three singular points. Two of
them are regular singularities situated at the origin and at the
horizon, $r=2\alpha$. The third one is at infinity and is an
irregular singularity of rank 2. Expansions of the solutions
around the singular points provide the boundary conditions for the
differential equations. The physical solutions are then selected
from the correct boundaries at the singularities. Gathering the
complete behaviour of the bound-state solution around the singular
points, we obtain the general form
\begin{equation}
\psi(r)= R(r) e^{-\sqrt{1-E^2}r+2i\sqrt{2\alpha\,r}E} \,
r^{\frac{\alpha(-1+2E^2)}{\sqrt{1-E^2}}} \, r^{-3/4},
\label{genral_sol}
\end{equation}
where $R(r)$ is an analytic function everywhere. The exponential
factor and the first term in $r$ take care of the singularity at
infinity \cite{cas}.  The $r^{-3/4}$ term corresponds to the boundary
at the origin.  The presence of this term is crucial as it accounts
for the non-Hermiticity of the Hamiltonian \cite{las2}. This can be
seen more explicitly from the calculation of the matrix elements of
${\bf h11}$ and ${\bf h22}$.  We find, for example, that
\begin{align}
 {\bf h11}_{lj}  
&= \int_{0}^{\infty} g_l^*\,g_j\,r^2\,dr +
   \sqrt{2 \alpha} \int_{0}^{\infty} g_l^*\, r^{3/4} i
   \partial_r(r^{3/4} g_j)] \, 
 dr \nn \\
&= \left[\int_{0}^{\infty} g_j^*\, g_l r^2 dr\right]^* +
   \sqrt{2 \alpha} \left[\int_{0}^{\infty} g_j^*\, r^{3/4} i
     \partial_r(r^{3/4} g_l^*) \,  dr\right]^* \nn \\
& \quad + i \sqrt{2 \alpha} r^{3/2} g_l^* g_j|_0^\infty \nn \\
&= {\bf h11^*}_{jl}+i \sqrt{2 \alpha} r^{3/2} g_l^*
   g_j|_0^\infty.
\label{nonh}
\end{align}
The second and third terms of the second equation follow from an
integration by parts. A similar calculation holds for ${\bf
h22}_{lj}$. As a consequence, if there is at least one pair $g_i$
and $g_j$ where both tend to the origin as $r^{-3/4}$, the
Hamiltonian is non-hermitian and the energy complex. The imaginary
part of the energy is always negative and represents the decay
rate of the wavefunction with time.

\section{Minimax for the black hole interaction}

Despite the fact that the interaction Hamiltonian is non-Hermitian, we
can still apply the minimax approach, if we select a suitable function
basis. The choice is based on the assumption that we achieve the
non-relativistic limit for low $\alpha$, in analogy with the Hydrogen
atom. Lasenby \emph{et al.} \cite{las} show that the non-relativistic
limits of the black hole and Coulomb interactions coincide. Therefore,
we expect that the energy spectrum for the black hole resembles the
Hydrogen spectrum for a low coupling constant. Restricting ourselves
to the low coupling regime, the task is then to choose a finite
function set that provides the Hydrogen properties and, at the same
time, has the correct boundaries at the singularities.

From equation (\ref{genral_sol}) we see that the bound state solutions
of the Dirac equation must go as $r^{-3/4}$ for $r\rightarrow 0$, and
as $e^{-br+i \beta \sqrt{r}}$ for $r\rightarrow \infty$.  Taking these
into
account we define a set of discrete basis
functions by
\begin{equation}
g_0=e^{i\,\sqrt{2\alpha r}} e^{-br} r^{-\frac{3}{4}}
\qquad 
g_j=e^{i\,\sqrt{2\alpha r}} e^{-br} r^{j-1} 
\label{funbasis1} 
\end{equation}
and
\begin{equation}
f_0 = e^{i\,\sqrt{2 \alpha r}} e^{-ar}r^{-\frac{3}{4}} 
\qquad
f_j = e^{i\,\sqrt{2\alpha r}} e^{-ar} r^{j-1} 
\label{funbasis2}
\end{equation}
where $j=1...n$, and $a$ and $b$ are additional (nonlinear)
parameters.  We limit ourselves for the case where $a$ and $b$ are
real. This corresponds to the condition $\Im(E)<<\Re(E)$, expected for
the non-relativistic limit. Note that we have increased the order of
the approximation to $n+1$ by including the extra terms $g_0$ and
$f_0$. From equation (\ref{nonh}), the matrix ${\bf h}$ will
non-hermitian if we introduce $g_0$ and $f_0$, and we find the the
non-hermitian terms of ${\bf h}$ from the quantities
\begin{equation}
\frac{\int_0^\infty g_0^*\, H_1(r) g_0\, r^2\, dr}{\int_0^\infty
  g_0^*\, g_0\, r^2\, dr} = -\alpha b-4i
  \sqrt{\frac{\alpha}{\pi}}b^\frac{3}{2}
\end{equation}
and
\begin{equation}
\frac{\int_0^\infty f_0^*\, H_1(r) f_0\, r^2\, dr}{\int_0^\infty
  f_0^*\,  f_0\, r^2\, dr} =
-\alpha a - 4i
       \sqrt{\frac{\alpha}{\pi}}a^\frac{3}{2} .
\label{expecH_1}
\end{equation} 
For the remaining terms down the diagonal of ${\bf h}$ we find that
\begin{equation}
\int_0^\infty g_j^* H_1(r) g_j\, r^2\,dr = -\int_0^\infty
g_j^*\left(\frac{\alpha}{r}\right) g_j\, r^2\,dr
\label{expE}
\end{equation}
with an identical result holding for $f_j$.  We can see here the
Coulomb interaction explicitly appearing in the diagonal of ${\bf h}$.

Leaving out the $g_0$ and $f_0$ terms in the expansion, the minimax
method reduces to a real energy approximation.  Thus, applying the
minimax method to a pair of functions $g_j$ and $f_j$ with $(j \neq
0)$ allows us to fix the values of $a$ and $b$. With these values, the
full basis can be determined, including the case $j=0$.  Our central
assumption now is that equation (\ref{eqn_genevalue}) is remains
valid, even when using the full basis set with $\bf{h}$ non-Hermitian.
The generalized eigenvalue problem of equation~(\ref{eqn_genevalue})
allows us then to calculate complex eigen-energies.  The validity of
our assumption rests in part on the fact that the results obtained are
in good agreement with the results of other, more accurate techniques.
Since equation (\ref{eqn_genevalue}) can be derived from a variational
principle of a Hermitian Hamiltonian, its validity for a non-Hermitian
Hamiltonian suggests the existence of a variational principle for such
a case.

\section{Results}

\begin{figure}
\begin{center}
\includegraphics[height = 6.5cm, width= 6.5cm ]{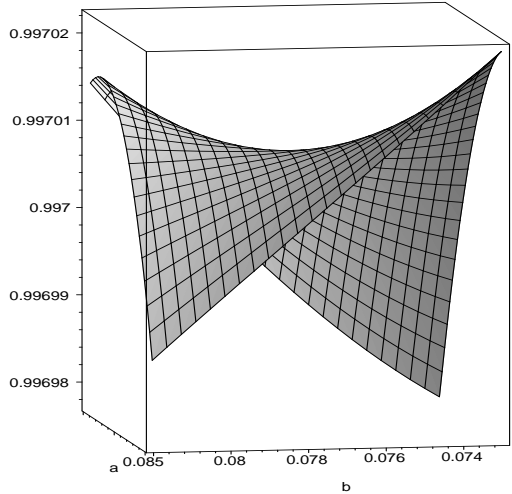}
\end{center}
\caption{\emph{Rayleigh quotient as function of the non-linear
parameters a,b}. $\alpha=0.1,\, k=-1$; minimax at: $
a=8.9186905\times 10^{-2},\, b=7.8238289\times 10^{-2}$.}
\label{Rcoeff}
\end{figure}

We start with the normalized basis function
\begin{equation}
\phi_*= 
\begin{pmatrix}
2 b^\frac{3}{2} e^{i\,\sqrt{2\alpha r}} e^{-br}\\
2 a^\frac{3}{2} e^{i\,\sqrt{2\alpha r}} e^{-ar}
\end{pmatrix},
 \label{BHtrial}
\end{equation}
which sets $j=1$ in equations (\ref{funbasis1}) and
(\ref{funbasis2}). As discussed, the Rayleigh quotient for this
function will be real. So the solution of equation (\ref{minmax}) is
given by the saddle point of $E_\phi$ as a function of $a$ and $b$. An
example of this saddle point is shown in figure~\ref{Rcoeff}.  From
equation (\ref{expE}) we see that the same result would have been
obtained for this function basis had we used the the Coulomb
interaction instead.

\begin{figure}
\begin{center}
\includegraphics[height = 8cm, width= 8cm, angle=-90 ]{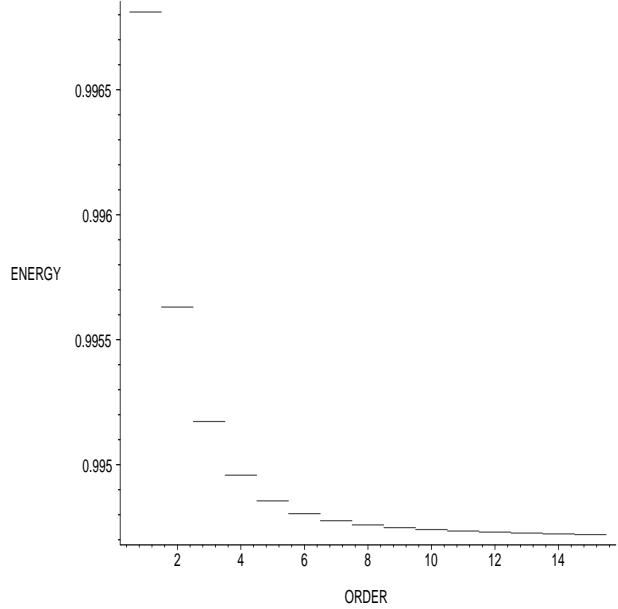}
\end{center}
 \caption{\emph{Convergence to a single value of the approximation to
 the real part of the energy $\Re(E)$}.  For values $\alpha=0.1,\,
 k=-1$, the energy $E/(mc^2)=0.9947208-i\,2.0498160 \times 10^{-5}$ is
 obtained.}
 \label{or_en_ground}
\end{figure}

The values of $a$ and $b$ determine a unique function set specified by 
equations (\ref{funbasis1}) and (\ref{funbasis2}). A suitable
trial function expanded on this function set is then
\begin{equation}
\phi_1= e^{i\sqrt{2\alpha r}}
\begin{pmatrix}
e^{-br} \left( b^0 r^{-3/4}+ \sum_{j=1}^{n}{b^{j}r^{j-1}} \right)
\\
e^{-ar} \left( a^0 r^{-3/4}+ \sum_{j=1}^{n}{a^{j}r^{j-1}} \right)
\end{pmatrix}.
\label{fulltrial}
\end{equation}
With this trial function, the eigenvalues of equation
(\ref{eqn_genevalue}) can be obtained. Since the function with power
$-3/4$ in equation (\ref{fulltrial}) gives rise to the non-Hermitian
terms in ${\bf h}$, the eigenvalues are complex. The approximation to
the ground state energy is then obtained from the eigenvalue with
minimum positive real part. As the order of the expansion is
increased, we expect the approximation to improve its accuracy and, in
particular, tend to a limit. That this is indeed the case is confirmed
in figure~\ref{or_en_ground}.

The minimax procedure offers only an approximation to the exact
bound-state energies. The accuracy of the method depends on the chosen
trial function. We consider two extra trial functions and compare
their results with a numerical shooting method that provides accurate
values of the ground-state energy~\cite{las}. The shooting method
finds the complex energies values for which the integrated solutions
from the boundaries at the horizon and at infinity coincide. For the
first trial function we consider the basis functions increasing on
$1/4$-rs powers of $r$. Then the trial function can be written as
\begin{equation}
\phi_2= e^{i\sqrt{2\alpha r}} 
\begin{pmatrix}
e^{-br} \left(\sum_{j=0}^{n}{b^{j}r^{-\frac{3}{4}+\frac{j}{4}}}
\right)
\\
e^{-ar} \left(\sum_{j=0}^{n}{a^{j}r^{-\frac{3}{4}+\frac{j}{4}}}
\right)
\end{pmatrix}.
\label{fulltrial14}
\end{equation}
For the second trial function we consider a combination of the trial
functions in equations (\ref{fulltrial}) and (\ref{fulltrial14}). The
new trial function increases in $1/4$ powers for the negative powers
of $r$ and in integer values for positive powers. This is
\begin{equation}
\phi_3= e^{i\sqrt{2\alpha r}} 
\begin{pmatrix}
e^{-br} \left( \beta^0 r^{-\frac{3}{4}}+\beta^1
r^{-\frac{1}{2}}+\beta^3 r^{-\frac{1}{4}}+
\sum_{j=0}^{n}{b^{j}r^{\frac{j}{4}}} \right)
\\
e^{-ar} \left( \alpha^0 r^{-\frac{3}{4}}+\alpha^1
r^{-\frac{1}{2}}+\alpha^3
r^{-\frac{1}{4}}+\sum_{j=0}^{n}{a^{j}r^{\frac{j}{4}}} \right)
\end{pmatrix}.
\label{fulltrialcomb}
\end{equation}
For $\alpha=0.1,\, k=-1$, we find the ground-state energies given in
table~\ref{table}.  (Note that the present conventions have $k$ with
opposite sign to the convention for $\kappa$ in~\cite{las}, so $k=-1$
is the ground state.)  The table shows that the approximation to the
real part of the complex energy given by the second trial function
($n=16$) is extremely good, with the imaginary part also obtained
quite accurately.  The third trial function ($n=18$) also gets the
real part quite accurately, but the decay rate approximation is poor.

\begin{table}
\centering
$\begin{array}{l|l}
             &  E/(mc^2) \\ \hline
$Trial function - minimax$\,\, \phi_1 & 0.9947208-
i\,2.0498160\times 10^{-5} \\
$Trial function - minimax$\,\, \phi_2 & 0.9946883-
i\,2.7855588\times 10^{-5} \\
$Trial function - minimax$\,\, \phi_3 & 0.9946858-
i\,3.0659648\times 10^{-5} \\
$Shooting method$ & 0.9946882- i\,2.7870824\times 10^{-5}
\end{array} $
\caption{\emph{Ground-state energy}. The energies are obtained
from the minimax, using different trial functions, and from the
shooting method.
\bigskip
} \label{table}
\end{table}

\begin{figure}[p]
\centering
\begin{minipage}{10cm}
  \centering
  \setlength{\unitlength}{1cm}
  \begin{picture}(10,7)
  \put(1.5,0.2){\makebox(7,7){\rotatebox{-90}{\includegraphics[height = 6.5cm, width= 6.5cm ]
  {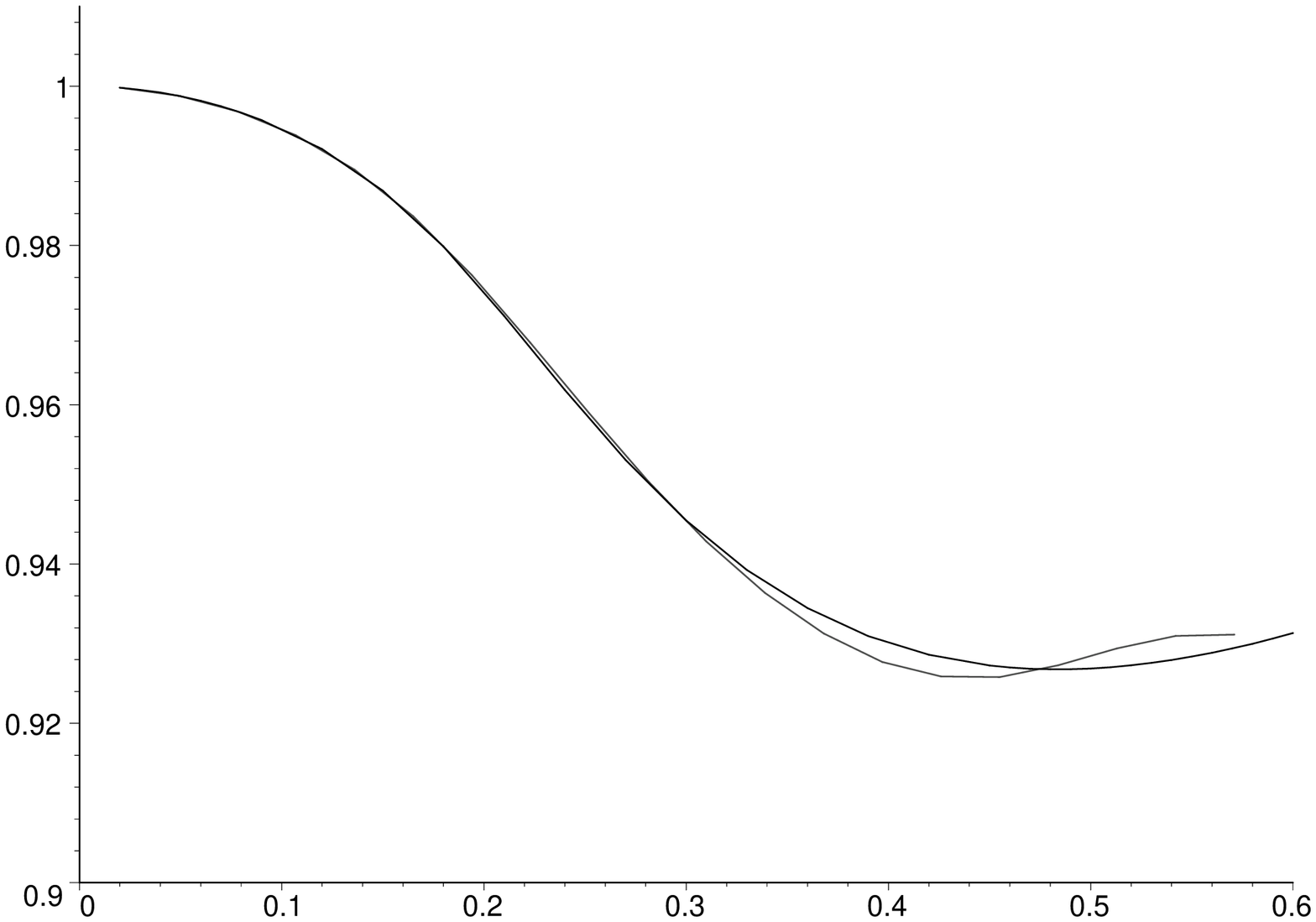}}}}
  \put(1,3.5){\makebox(0.5,0.5){\footnotesize{$\Re(E)/(mc^2)$}}}
  \put(5,0) {\makebox(0.5,0){\footnotesize{$\alpha$}}}
  \end{picture}
 \end{minipage}
 \caption{\emph{Real part of the energy as a function of the black hole
mass}. The real part of the energy given by the minimax (light
hue) and by the shooting energy (dark hue) are compared for the
ground state with $k=-1$.}
 \label{shoot E}
\end{figure}

\begin{figure}[p]
\centering
\begin{minipage}{10cm}
  \centering
  \setlength{\unitlength}{1cm}
  \begin{picture}(10,7)
  \put(1.5,0.2){\makebox(7,7){\rotatebox{-90}{\includegraphics[height = 6.5cm, width= 6.5cm ]
  {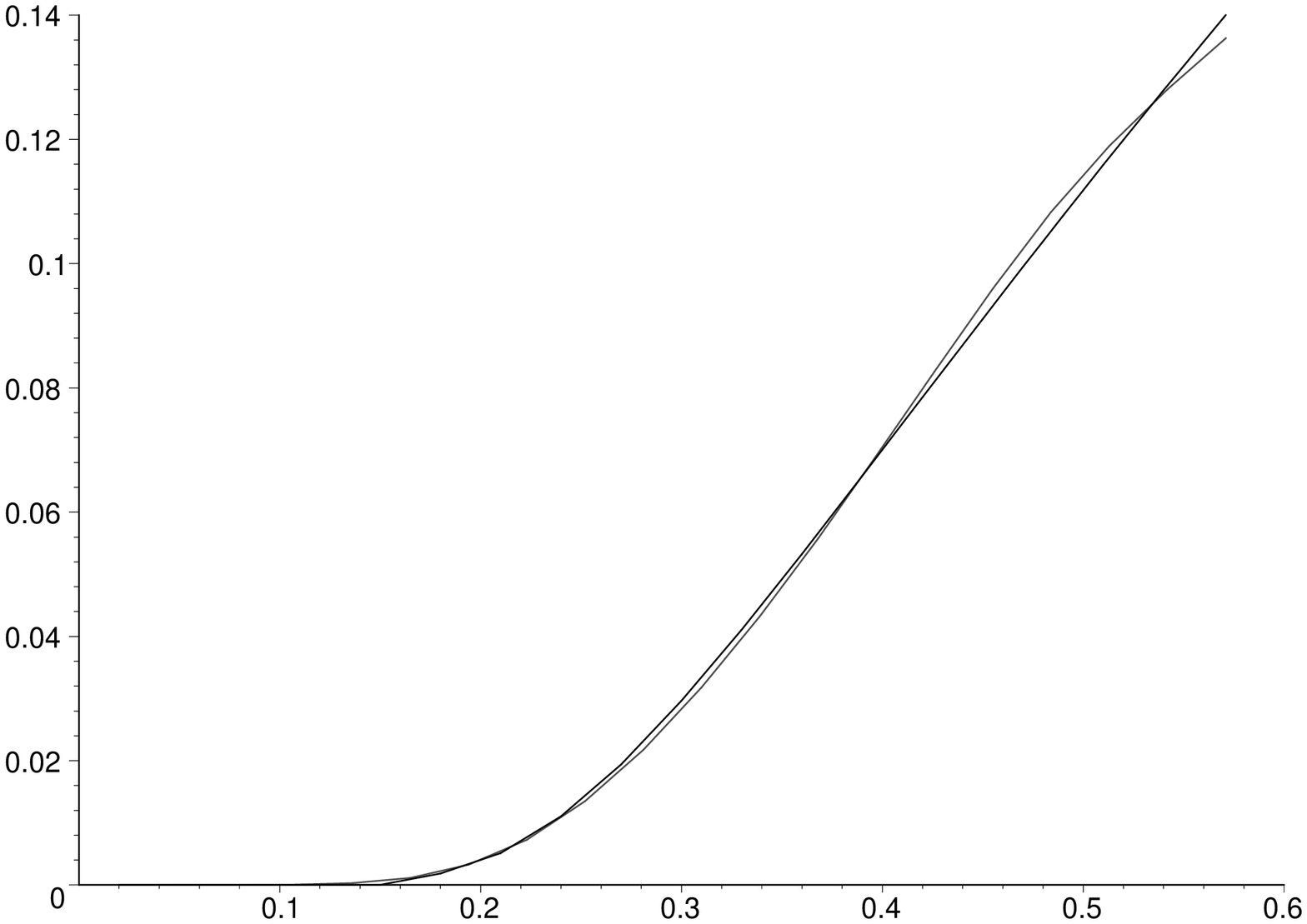}}}}
  \put(1,3.5){\makebox(0.5,0.5){\footnotesize{$\Im(E)/(mc^2)$}}}
  \put(5,0) {\makebox(0.5,0){\footnotesize{$\alpha$}}}
  \end{picture}
 \end{minipage}
 \caption{\emph{Decay rate as a function of the black hole
mass}. The decay rates given by the minimax (light hue) and the
shooting (dark hue) corresponding to figure~\ref{shoot E}.}
 \label{shoot ep}
\end{figure}

The trial function $\phi_2$ is then more suitable for the study of the
behaviour of the method with an increasing coupling constant.  This is
shown in figures~\ref{shoot E} and~\ref{shoot ep}, where we can see
that the minimax method gives good results for low couplings. Around
the point of the last stable circular orbit $\alpha=0.28$ \cite{las}
the minimax loses accuracy. This gives numerical indication that the
region where the minimax works at its best is where classical circular
orbits are still admissible.

\begin{figure}[p]
\centering
   \begin{minipage}{10cm}
     \centering
     \setlength{\unitlength}{1cm}
      \begin{picture}(10,8)
      \put(0,0){\makebox(10,8){
        \rotatebox{-90}{
        \includegraphics[height = 7.5cm,width=
        7.5cm]{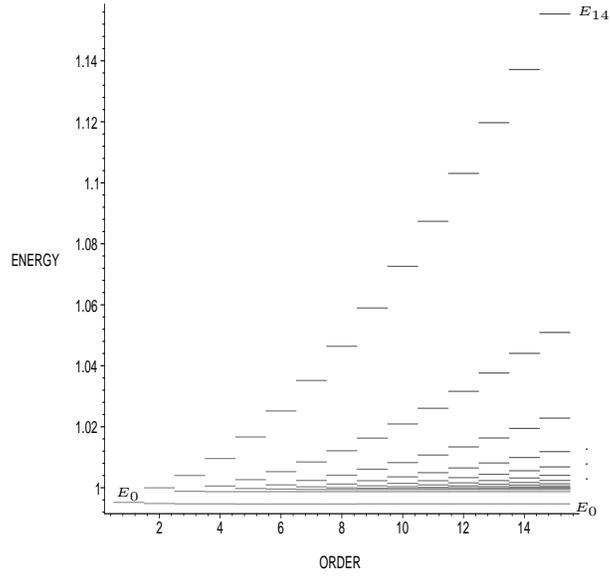}}
        }}
       \put(2.5,0.9){\makebox(0.5,0.5){\tiny{$E_0$}}}
       \put(8.7,7.3){\makebox(0.5,0.5){\tiny{$E_{14}$}}}
       \put(8.6,0.7){\makebox(0.5,0.5){\tiny{$E_{0}$}}}
       \put(8.6,1.1){\makebox(0.5,0.5){\tiny{$.$}}}
       \put(8.6,1.3){\makebox(0.5,0.5){\tiny{$.$}}}
       \put(8.6,1.5){\makebox(0.5,0.5){\tiny{$.$}}}
      \end{picture}
   \end{minipage}
\caption{\emph{Hylleraas--Undheim behaviour of the positive real
part of the eigen-energies}, given the values $\alpha=0.1,\, k=-1$
and the trial function (\ref{fulltrialcomb}).} \label{hyll_und1}
\end{figure}

\begin{figure}[p]
\centering
   \begin{minipage}{10cm}
     \centering
     \setlength{\unitlength}{1cm}
      \begin{picture}(10,8)
      \put(0,0){\makebox(10,8){
        \rotatebox{-90}{
        \includegraphics[height = 7.5cm, width= 7.5cm ]
        {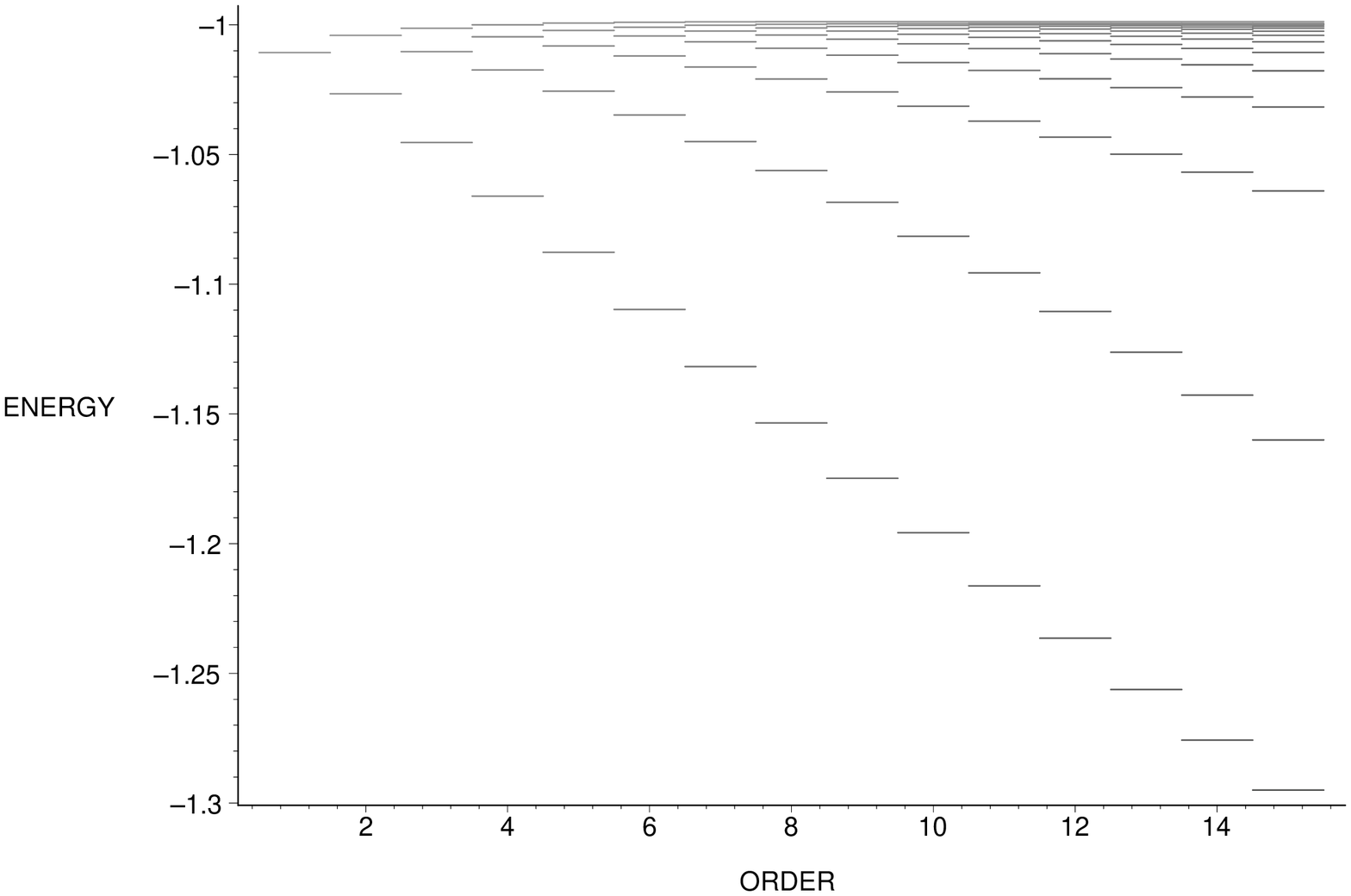}}}
        }
      \end{picture}
   \end{minipage}
\caption{\emph{Hylleraas--Undheim of the negative real part of the
eigen-energies}, given the values $\alpha=0.1,\, k=-1$ and the
trial function (\ref{fulltrialcomb}).} \label{hyll_und2}
\end{figure}

The minimax method also provides approximations for the excited states.
These are obtained from the other real positive eigenvalues. Using the
trial function $\phi_3$ a neat Hylleraas--Undheim behaviour is obtained
for $\Re(E)$. This is shown in figures~\ref{hyll_und1}
and~\ref{hyll_und2} where the real eigenstates tend to different
energy levels as the order increases. In contrast to the hydrogen
case, we have not only positive energy but also negative energy
states. A selection of the values for the bound energies obtained in
these figures are given in table~\ref{energy_table}, where the first
two eigenvalues of positive and negative energies are compared for the
cases $k=-1$ and $k=1$, with $\alpha=0.1$.

\begin{table}[b]
$\begin{array}{ll|l|l}

        &        & $Positive Energy$/mc^2
                 & $Negative Energy$/mc^2                \\ \hline
   k=-1 & 1^{st} & 0.994686-i\,3.06596 \times 10^{-5}
                 & -0.998730-i\,1.53600 \times 10^{-8}\\
        & 2^{nd} &  0.998698-i\,3.83240 \times 10^{-6}
                 & -0.999437-i\,4.74152 \times 10^{-7}\\
   k=1  & 1^{st} &  0.998731-i\,2.00712 \times 10^{-7}
                 & -0.985823-i\,1.55033 \times 10^{-2}\\
        & 2^{nd} &  0.999438-i\,5.22900 \times 10^{-8}
                 & -0.994978-i\,1.75747 \times 10^{-3}
\end{array}$
\caption{\emph{Bound-state Energies for excited
states}}
\label{energy_table}
\end{table}

From this table we can extract a number of features.  There is a
spurious root for the first negative state with $k=1$. All the other
energies can be checked in a numerical integration routine and
confirmed to give bound states. The solution for this spurious root
does not show any structure other than an exponential divergence from
the origin. The presence of spurious roots in variational methods for
the Dirac equation is not new. In the past these roots have appeared
from the study of the Coulomb interaction. In previous works, Drake
and Goldman \cite{gol,gol2} found the presence of a spurious root for
the first eigenvalue of the case $k=1$. In his original
work~\cite{talman} Talman showed how, for a particular case, the
minimax method removes the spurious root found in \cite{gol}. Note
that this kind of root is not present in table~\ref{energy_table}, so
the minimax procedure has certainly excluded this case and given a
sensible value for the first positive eigenvalue with $k=1$. The
spurious root is found, however, for the first negative eigenvalue of
$k=1$. This lies outside the scope of Talman's arguments, since
negative energy states of this type are not present in Coulomb
interactions.

Another feature of table~\ref{energy_table} is given by the comparison
of the energies of the states $2S_{1/2}$ $(2^{nd},k=-1)$ and
$2P_{1/2}$ $(1^{st},k=1)$. For the low value of $\alpha$ used here, we
see that the states have comparable energy, as is the case for the
Coulomb interaction. At higher values of $\alpha$ for the
gravitational interaction this degeneracy of the energy in the sign of
$k$ is lifted~\cite{las}.

A further property we can extract from table~\ref{energy_table} is the
presence of charge conjugation symmetry.  For a given state we should
find another with opposite sign for both the real part of the energy
and $k$, but with the same decay rate. This can be seen from the
comparison between the real part of the state $(1^{st}, k=-1)$ of
positive energy and the state $(2^{nd},k=1)$ of negative energy, as
well as the comparison between the states of positive energy
$(1^{st},k=1)$, $(2^{nd},k=1)$ and the states of negative energy
$(1^{st},k=-1)$, $(2^{nd},k=-1)$ respectively.  It is a bit
disappointing that the imaginary energies of corresponding states have
not the same value, as is required by the charge conjugation symmetry.
This suggest that the minimax method will need to be further refined
to find a more precise decay rates for the excited states.

\section{Conclusions}

The minimax method has proven to be a useful tool for analyzing the
electron bound state energy spectrum in a black hole background. We
have seen how the method works well not only for the small coupling
limit but also sheds some light on what happens beyond the last
classically stable circular orbit. The validity of the method suggests
that the complex energy spectrum of the black hole can be obtained
from a general variational principle.

The minimax method can only give an approximation to the exact
energy. As we have seen this depends critically on the trial function
used.  Furthermore, we found that the method ceases to give reliable
results around $\alpha \sim 0.4$, which may limit the applicability of
the minimax method. One reason for this failure may be that we have
assumed that the imaginary part of the energy is small as compared to
the real part.  This is no longer the case around $\alpha \sim 0.4$
for the $1S$ state.  However, we may expect that the minimax method
will continue to give accurate results for higher angular momentum
states beyond $\alpha \sim 0.4$.

A further improvement of the minimax method can be given by the
relaxation of the condition on the imaginary energy. As a consequence,
the nonlinear parameters $a$ and $b$ could be considered as complex
numbers.  It then would be interesting to see whether this gives any
improvements in either the determination of the energy for higher
values of the black hole mass, or to the accuracy of the decay rates
for excited states.  Extending this method to larger values of
$\alpha$ is a particularly significant problem, as the shooting method
runs into numerical difficulties beyond $\alpha$ of around 6 , whereas
values of astrophysical interest start at around $10^{15}$.


\begin{thebibliography}{10}

\bibitem{talman} 
J.D.~Talman.
\newblock Minimax principle for the Dirac equation.
\newblock \emph{Phys. Rev. Lett.} {\bf 57}:1091, 1986.

\bibitem{hylleraas} 
E.A.~Hylleraas and B.~Undheim.
\newblock Numerische berechuwny der 2S-Terme von ortho-und par-helium. 
\newblock \emph{Z. Phys} {\bf 65}:759, 1930.

\bibitem{mac} 
J.K.L.~MacDonald.
\newblock Successive approximations by the Rayleigh--Ritz variational
method.  
\newblock \emph{Phys. Rev.} {\bf 43}: 830, 1933.

\bibitem{sch} 
W.H.~Schwarz and H.~Wallmeier.
\newblock Basis set expansions of relativistic molecular wave
equations.  
\newblock \emph{Mol. Phys.} {\bf 46}:1945, 1982.

\bibitem{dol} 
J.~Dolbeault, M.J.~Esteban and E.~Sere. 
\newblock Variational characterisation for eigenvalues of Dirac
operators.  
\newblock \emph{Cal. Var.} {\bf 10}: 321, 1998.

\bibitem{dol2} 
J.~Dolbeault, M.J.~Esteban, E.~Sere and M.~Vanbrengel.
\newblock Minimisation methods for the one-particle Dirac equation.
\newblock \emph{Phys. Rev. Lett.} {\bf 85}:4020, 2000.

\bibitem{gol} 
S.P.~Goldman and G.W.F.~Drake.
\newblock Application of discrete-basis-set methods to the Dirac
equation. 
\newblock \emph{Phys. Rev. A} {\bf 23}:2093, 1981.
 
\bibitem{grant} 
I.P.~Grant.
\newblock Variational methods for Dirac wave equations.
\newblock \emph{J. Phys. B.} {\bf 19}: 3187, 1986.

\bibitem{gol2} 
S.P.~Goldman.
\newblock Variational representation of the Dirac--Coulomb Hamiltonian
with no spurious roots. 
\newblock \emph{Phys. Rev. A.} {\bf 31}:3541, 1985.

\bibitem{las} 
A.N.~Lasenby, C.J.L.~Doran, J.~Pritchard, A.~Caceres and S.~Dolan.
\newblock Bound states and decay times of fermions in a
Schwarzschild black hole background. 
\newblock gr-qc/0209090, 2002.

\bibitem{las2} 
A.N.~Lasenby, C.J.L.~Doran and S.F.~Gull.
\newblock Gravity, gauge theories and geometric algebra.
\newblock \emph{Phil. Trans. R. Soc. Lond. A} {\bf 356}: 487, 1998.

\bibitem{cas} 
A.~Cacares.
\newblock \emph{Electron Dynamics on a Black Hole Background}.
\newblock Ph.D. Thesis, University of Cambridge, 2004.

\bibitem{las3} 
A.N.~Lasenby and C.J.L.~Doran.
\newblock Geometric algebra, wave functions and black holes. 
\newblock \emph{Advances in the Interplay between Quantum and Gravity
  Physics,} Kluwer, pg: 251-283, 2002. 

\bibitem{brill} 
D.R.~Brill and J.A.~Wheeler.
\newblock Interaction of neutrinos and gravitational fields.
\newblock \emph{Revs. Mod. Phys.} {\bf 29}:465, 1957.

\bibitem{bj} 
J.D.~Bjorken and S.D.~Drell.
\newblock \emph{Relativistic Quantum Mechanics, Volume I.} 
\newblock McGraw-Hill, 1964.

\bibitem{dor} 
C.J.L.~Doran and A.N.~Lasenby.
\newblock \emph{Geometric Algebra for Physicists.}
\newblock Cambridge University Press, 2003.

\bibitem{gr} 
W.T.~Grandy.
\newblock \emph{Relativistic Quantum Mechanics of Leptons and Fields.} 
\newblock Kluwer Academic, Dordrecht, 1994 .

\bibitem{gri} 
M.~Griesemer, R.T.~Lewis and H.~Siedentop. 
\newblock A minimax principle for eigenvalues in spectral gaps: Dirac
operators in Coulomb potentials. 
\newblock \emph{Doc. Math} {\bf 4}: 275, 1999. 

\bibitem{Ter} 
I.M.~Ternov and A.B.~Gaina, A. B. 
\newblock Energy spectrum of the Dirac equation for the Schwarzschild
and Kerr fields 
\newblock \emph{Sov. Phys. J.} {\bf 31}(2):157, 1988.

\end{thebibliography}
\end{document}